\documentclass[twocolumn,prd]{revtex4}

\usepackage{graphicx}
\usepackage{dcolumn}
\usepackage{bm}

\newcommand{\beq}{\begin{equation}}
\newcommand{\eeq}{\end{equation}}
\newcommand{\beqa}{\begin{eqnarray}}
\newcommand{\eeqa}{\end{eqnarray}}

\def\l({\left(}
\def\r){\right)}

\begin{document}

\title{Which blazars are  neutrino loud?}

\author{A.Yu.~Neronov$^a$ and  D.V.~Semikoz$^{bc}$}
\affiliation{$^a$ Theoretische Physik, Universit\"at M\"unchen, Theresienstr. 37, 
80333 Munich, Germany,\\
$^b$ Max-Planck-Institut f\"ur Physik,
F\"ohringer Ring 6, 80805 Munich, Germany\\
$^c$ Institute for Nuclear Research, 60th October Anniversary, 7a,
Moscow, 117312, Russia}

\begin{abstract}
Protons accelerated in the cores of active galactic nuclei can effectively produce 
neutrinos only if the soft radiation background in the core is sufficiently high.  We find restrictions 
on the spectral properties and luminosity of blazars under which they can be strong neutrino sources. 
We analyze the possibility that neutrino flux is highly beamed along the 
rotation axis of the central black hole. The enhancement of neutrino flux 
compared to GeV $\gamma$-ray flux from a given source makes the detection of neutrino point sources 
more probable. At the same time the smaller open angle reduces the number of possible neutrino-loud blazars 
compared to the number of $\gamma$-ray loud ones. 
We present the table of 15 blazars which are the most likely candidates for the detection by future neutrino telescopes. 
\end{abstract}

\maketitle

\section{Introduction}

Neutrino telescopes which already operate or under construction will presumably be able to 
detect point sources of neutrinos with energies  up to
$\sim10^{17}\,$eV by looking for the showers and/or tracks
from charged leptons produced by charged current reactions of
neutrinos in ice, in the case of AMANDA~\cite{amanda,amanda_limit}
and its next generation version ICECUBE~\cite{icecube},
in water, in the case of BAIKAL \cite{baikal,baikal_limit}, 
ANTARES~\cite{antares}, and NESTOR ~\cite{nestor}
(for recent reviews of neutrino
telescopes see Ref.~\cite{nu_tele}).

From the other side future Ultra-High Energy Cosmic Ray (UHECR) 
experiments like the Pierre Auger Observatory~\cite{auger,auger_nu} 
will be able to detect neutrinos with energies above $10^{17}\,$eV. 
At the highest energies above $\sim10^{19}\,$eV the  telescope array~\cite{ta,ta_nu} or 
 space based observatories such as EUSO~\cite{euso} and OWL~\cite{owl,owl_nu}
will also measure the neutrino flux. The flux from point-like sources at those 
energies can be a combination of  the direct flux from a source and the secondary neutrino flux
produced by Ultra-High Energy (UHE) protons emitted by the source in interactions with cosmic microwave background radiation (CMBR) photons.
If neutrino flux at high energies $E>10^{17}$ eV is large enough
to give more then 1-3 events per km$^2$ per year, future km$^2$
neutrino telescopes like ICECUBE will be able to detect
those neutrinos coming from above \cite{halzen-alvarez}.

At the intermediate energies between $\sim10^{15}\,$eV and $\sim10^{19}\,$eV 
there are plans to construct telescopes to detect
fluorescence and \v{C}erenkov light from near-horizontal showers produced in
mountain targets by neutrinos ~\cite{fargion,mount}.
The alternative of detecting neutrinos by triggering onto the
radio pulses from neutrino-induced air showers is also currently 
investigated~\cite{radhep}. 
Two implementations of this technique,
RICE, a small array of radio antennas in the South pole ice~\cite{rice},
and the Goldstone Lunar Ultra-high energy neutrino Experiment (GLUE)~\cite{glue}, 
have so far produced neutrino flux upper limits.
Acoustic detection of neutrino induced interactions is also
being considered~\cite{acoustic}.

The simplest way to produce neutrinos in astrophysical objects is to accelerate protons 
and then collide them with soft photon background with energy above  photo-pion production 
threshold. The produced pions will decay in photons, electrons, positrons  and neutrinos. If protons
are captured within the source, the estimate on neutrino flux from a given 
source can be obtained from the detected $\gamma$-ray flux, since
the energy deposit in neutrinos in pion decays is of the same order as 
the energy in photons. 
If the sources are transparent for the primary protons than 
a limit on diffuse neutrino flux from all the possible sources 
can be obtained from the detected high-energy proton flux. 
This idea was first suggested in \cite{halzen}. 
For a particular case of $E^{-2}$ proton spectrum
coming from AGN's the calculation was done in \cite{wb}.
The same calculation for $E^{-1}$ proton flux was made in \cite{mpr}.
In \cite{diffuse} the dependence of neutrino flux from proton
spectrum, cosmological parameters and distribution of sources was
investigated in details.  In particular it was shown that in
many cases neutrino flux can exceed
value calculated in \cite{wb} or even value of 
\cite{mpr} and only bound on diffuse
neutrino flux come from EGRET measurement.

The Universe is not transparent for photons with energies above 100 GeV. The highest
energy photons from astrophysical objects (nearby TeV blazars) seen so far had energies 
$E \sim 10^{13}$ eV.  
No direct  information about emission  of $E > 10^{13}$ eV particles
is available now. 
At the same time  it is well established that photon 
emission from blazars (active galactic nuclei (AGN), which we see almost face on) in the MeV-TeV energy range is 
 highly anisotropic.
 Typical estimates of the 
$\gamma$ factors of the emitting plasma, $\gamma\sim 10$, imply that in the $10^{6-13}$ eV band
almost all $\gamma$-ray flux  is radiated 
in a cone with the opening angle $\theta\sim 1/\gamma\sim 5^\circ$. Particles (photons, neutrinos) 
in the higher energy range $E> 10^{13}$ eV can be emitted in an even narrower cone. 
This fact favors blazars  as promising neutrino sources.     

Recent X-ray observations of large-scale jets in AGN can shed some light on the issue of 
particle acceleration to the energies much above TeV in the AGN cores. Indeed, in order 
to explain X-ray synchrotron emission on very large scales of order of 100 kpc away from the 
AGN core one needs to suppose that multi-TeV electrons are continuously produced over the whole
jet length. A model which naturally explains this continuous production of multi-TeV electrons 
was recently proposed in \cite{gamma_jet} (see Fig. \ref{fig:cartoon} and \ref{fig:blazar_gev}). 
The idea is that $\gamma$-rays with energies $10^{14-16}$ eV
emitted from the  AGN core  produce $e^+e^-$ pairs in interactions with the CMBR photons  at the distance scale  10-100 kpc away from the core. 
 Thus, within this model the fact that jets with the lengths about 10-100 kpc are commonly 
observed in AGNs  enables to conclude that (1) particles with energies $ E \ge 10^{14-16}$ eV are produced in 
the AGN cores and (2) these particles are normally emitted in a cone with opening angle 
$\theta\sim 1^\circ$. 
  The diffuse neutrino flux in this model was calculated in \cite{diffuse}.
In this paper we  discuss which blazars will be the most promising 
neutrino sources if neutrinos are produced in the AGN cores, as in the model \cite{gamma_jet}. AGN which can be significant point sources of neutrinos 
were analyzed in \cite{eichler,stecker,atoyan,halzen}. In particular, 
high neutrino fluxes were conjectured to come from brightest quasars like
3C 273 \cite{eichler,stecker} or TeV blazars, like Mkn 421 \cite{halzen}. 
Enhancement of neutrino flux during the flaring activity in 3C 279 
was considered in \cite{atoyan}. Predictions of the model \cite{gamma_jet}
are quite different. In particular, none of the three
above cited blazars enters our list of most probable neutrino sources.

In Section  \ref{sec:two} we will discuss a mechanism of neutrino production and derive a 
bound on the magnitude and redshift of blazars which can be neutrino-loud.
In the third section we discuss the neutrino flux from TeV gamma-ray sources. 
In  Section \ref{sec:four} we will derive the neutrino fluxes from 15 blazars which are the most promising 
neutrino sources. In Section \ref{sec:five}  we will discuss the secondary neutrino fluxes from
UHECR sources.

\section{Production of neutrinos in blazars.}
\label{sec:two}

If an AGN is expected to be a bright neutrino source, physical conditions inside the AGN core must be 
favorable for intense production of neutrinos in photo-pion process. This means that the density of 
soft photons in the core must be high enough for protons to interact at least once with the  background 
photons while they traverse the core. Of course, the soft photon density in the core is expected to be 
highly anisotropic and the mean free path of protons depends essentially on the direction of 
propagation. The main contribution to the soft photon background in the direct vicinity of the 
central black hole is usually assumed to 
come from the optical/UV (or "blue bump") photons produced by the inner part of accretion disk.

\begin{figure}
\begin{center}
\includegraphics[width=0.9\linewidth,clip=true,angle=0]{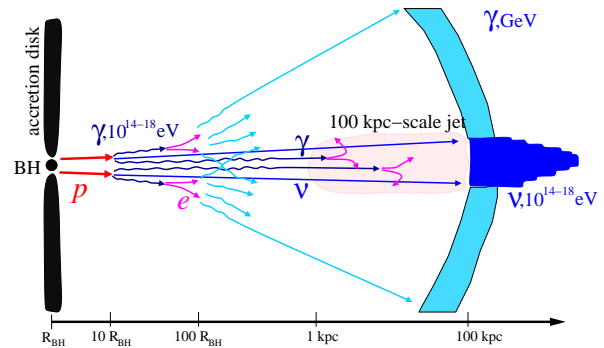}
\caption{Neutrino production in the AGN core.}
\label{fig:cartoon}
\end{center}
\end{figure}
\begin{figure}
\includegraphics[width=0.6\linewidth,clip=true,angle=270]{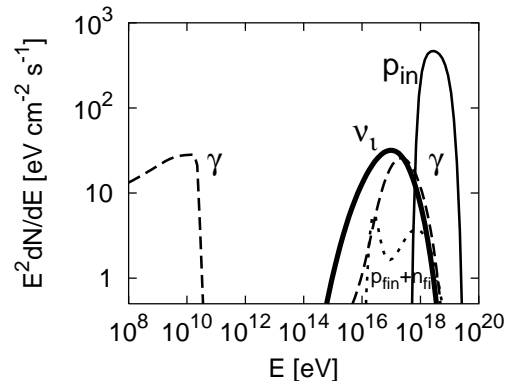}
\caption{Typical particle spectra of GeV-loud blazar in the model presented in Fig.\ref{fig:cartoon}.
The thin solid line is the initial spectrum of accelerated protons. Dotted line is the final spectrum
of protons and neutrons after interaction with UV photons. Thick solid line presented averaged
over flavor spectrum of neutrinos. Photons are shown by dashed line. High energy photons will feed 
a kpc-scale jet and partly  contribute to the GeV region. Spectra are normalized to the flux of a typical 
GeV-loud EGRET source.}    
\label{fig:blazar_gev}
\end{figure}

As an example, the neutrino production mechanism in model \cite{gamma_jet} is presented schematically in 
Fig.~ \ref{fig:cartoon}. A beam of high-energy protons accelerated in a strong electromagnetic field in the 
direct vicinity of the black hole horizon is converted in the core 
into a beam of secondary particles such as $\gamma$-quanta, neutrinos, electrons and positrons. 
The beam of 
high-energy $\gamma$-quanta feeds the bright 100-kpc scale jet with high-energy electrons, while the beam 
of neutrinos just escapes the source.  Typical particle spectra of GeV-loud blazar in this model are shown  in 
Fig. \ref{fig:blazar_gev}.  The spectra are calculated using the code developed in \cite{code}.

For AGNs which are not seen in TeV range, no direct estimate of optical depth for protons in a given 
direction is possible because $\gamma$-rays with energies below 100 GeV do not interact with 
the "blue bump" photons in the AGN core. 
In this case one can roughly estimate the conditions inside the AGN core 
assuming that the optical/UV background is isotropic and is produced inside the region of the 
size of order of $R_{\rm core}\sim 10^{16}$ cm as it is indicated by typical optical/UV variability of 
AGNs \cite{opt_var}.  
If the flux from the source is $\nu F_\nu$, the luminosity is
$L=4\pi D_L^2(\nu F_\nu)$
where 
$D_L$ is the luminosity distance. The number density of 
photons inside the core is
\beq
n_{\rm soft}=\frac{L}{4\pi R^2_{\rm core}c\epsilon}=\frac{D_L^2 (\nu F_\nu)}{R^2_{\rm core} c\epsilon}~,
\eeq 
where $\epsilon=\epsilon_V\times 2.26$ eV is the typical energy of soft photons in the V-band.
The mean free path of the proton is given by
\beq
R_p=\frac{1}{\sigma_{p\gamma} n_{\rm soft}}=\frac{cR_{\rm core}^2\epsilon}{\sigma^{p\gamma}(\nu F_\nu) D_L^2}~.
\eeq
The requirement that the proton mean free path  is much smaller than $R_{\rm core}$ imposes the 
restriction on the flux in the V band from blazars at a given redshift,
\beq
\label{bound}
\frac{D_{L,100}^2(\nu F_{\nu})_{\rm V, 13} \sigma^{p\gamma}_{28}}{\epsilon_V R_{16}}\gg 1~,
\eeq
where $D_{L,100}$ is the luminosity distance in units of 100 Mpc, size of the core $R_{16}=R/(10^{16} {\rm cm})$,
normalized cross section $\sigma^{p\gamma}_{28}=\sigma^{p\gamma}/10^{-28} {\rm cm}^2$ and the  flux normalized on 
magnitude $m_V=13$ is
$(\nu F_\nu)_{\rm V}=1.26 \times 10^{-13} (\nu F_{\nu})_{\rm V, 13}  {\rm W}/{\rm m}^2$. Here
 $(\nu F_{\nu})_{\rm V, 13}=10^{-0.4 (m_V-13)}$.

The luminosity distance depends on the cosmological model. For our calculations we choose 
the best motivated at present cosmological model. Namely, a  flat Universe, 
filled with matter  $\Omega_M = \rho_M/\rho_c$ 
and vacuum energy densities $\Omega_V=\rho_V/\rho_c$ whose sum equals the critical energy density,
 $\Omega_V + \Omega_M = 1$. The critical energy density $\rho_c=3H_0^2/(8\pi G_N)$ is defined through 
the Hubble parameter $H_0$.
The luminosity distance in this model has the following form: 
\beq
\label{Dl}
D_{L}=\frac{1+z}{H_0\sqrt{\Omega_M}} \int_1^{1+z} \frac{dx}{\sqrt{\frac{\Omega_V}{\Omega_M}+x^3}}~.
\eeq

In our calculations we used  the values $H_0=70 {\rm km}/{\rm s} /{\rm Mpc}$, $\Omega_V=0.7$  and $\Omega_M=0.3$.

 The bound Eq. ~(\ref{bound}) imposes a restriction on the blazar redshift and 
 magnitude which is shown on Fig. \ref{fig:L_Dl} (the solid line corresponds to the case  when the 
left hand  side 
of Eq. ~(\ref{bound}) equals 5. Such a
choice is motivated by the fact that in one photo-pion interaction  a proton 
looses only a fraction (typically 20\%) of its energy.

An estimate (but not an upper limit, see below Section \ref{sec:four}) 
of the  neutrino flux from sources which satisfy the constraint Eq. ~(\ref{bound}) can be obtained
from the detected $\gamma$-ray flux. The GeV $\gamma$-rays can be produced in the AGN core through a variety of mechanisms: inverse Compton scattering of soft background photons like in synchrotron-self Compton model \cite{ssc},
 synchrotron  radiation of very-high energy protons in  extreme proton synchrotron model
\cite{aharonian2000}, development of 
electromagnetic cascade initiated by photo-pion production in proton blazar models \cite{pic}. Neutrinos can be produced
only in the last case. In the following we will suppose that  photo-pion production gives a significant 
contribution to the observed GeV photon flux, allowing us to estimate the resulting neutrino flux.

\section{Can TeV-loud blazars be neutrino sources?}

Since the high energy neutrinos are produced together with the high energy photons in photo-pion 
reaction, the presence 
of high energy photons in the spectrum of a given source can serve as an indicator
for a possible neutrino flux from this source.
However, produced high energy photons still can interact with the background
photons both in the source and on the way to the Earth, cascading down
to the energies below the pair production threshold. The typical energy 
of the ``blue-bump'' photons in the source (1-10 eV) is similar to
the energy of infrared background photons. Thus, the high energy  photon spectrum 
should end  in the 10 GeV - TeV region, depending on the source properties
and the distance to the Earth. Thus, promising neutrino sources 
should be GeV or TeV loud.

For our analysis  we have taken the catalog of blazars which are detected 
by  EGRET \cite{blazars} (51 objects), singled out the ones which are listed as 
GeV sources   \cite{GeVEGRET}
(21 sources) and added all known TeV gamma-ray sources (8 sources) presented in the Table \ref{tab:tev}.
TeV gamma-rays from the first three sources in this table are confirmed by several experiments
with high significance, while the last 5 sources have been seen only by one experiment
and require confirmation in the future. Two of the sources,  Mkn 421 and 3C 66A, were also
seen by EGRET with significant flux in GeV region. Thus, the total number of
sources in our analysis is 27.  We used NED \cite{NED} and Simbad \cite{Simbad} databases 
to find V-magnitudes  and redshifts of the objects.

\begin{table}
\begin{center}
\begin{tabular}{|c|l|l|l|l|l|}
\hline
N & NAME   & ~~{\it z}   &V mag  & Type & Telescopes\\
\hline
1&Mkn 421  & 0.031 & 13.5 & HBL & CAT, HEGRA,   \\
&&&&&WHIPPLE \cite{Mkn421}\\
\hline
2&Mkn 501  & 0.0337 & 13.8 & HBL & CAT, HEGRA,   \\
&&&&&WHIPPLE \cite{Mkn501}\\
\hline
3&1ES 1426+428  & 0.129 & 16.5 & HBL & CAT, HEGRA,   \\
&&&&&WHIPPLE \cite{1ES1426+428}\\
\hline
4&1ES 2344+514  & 0.044 & 15.5 & HBL & WHIPPLE \cite{1ES2344+514}   \\
\hline
5&1ES 1959+650  & 0.047 & 14.7 & HBL & Utah Tel. Array \cite{1ES1959+650}   \\
\hline
6& BL Lac  & 0.0686 & 14.5 & LBL & Crimean Obs. \cite{BLLac}   \\
\hline
7& PKS 2155-304  & 0.117 & 13.1 & HBL & Durham Mark 6 \cite{PKS2155-304}   \\
\hline
8& 3C 66A  & 0.444 & 15.5 & LBL & Crimean Obs \cite{3C66A}   \\
\hline
\end{tabular}
\caption{TeV loud blazars. Number in the first column same as in Fig.~\ref{fig:L_Dl}.
Second column is the name of the object.  Third is redshift. Forth is magnitude in V range
(eV range). Fifth is BL Lac type, and the last is experiment name and reference.} 
\label{tab:tev}                             
\end{center}
\end{table}

\begin{figure}
\includegraphics[width=0.6\linewidth,clip=true,angle=270]{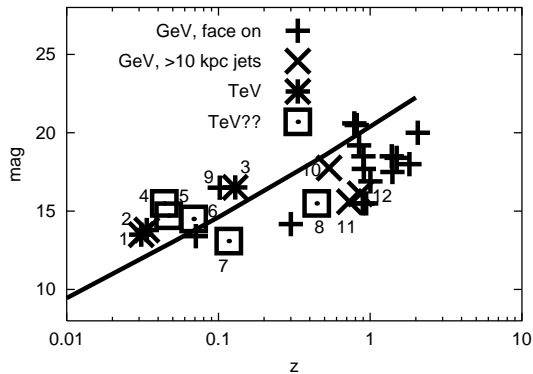}
\caption{Bound on the magnitude and redshift of blazars which can be neutrino-loud, Eq.~ (\ref{bound}). Above 
the solid line only part of proton energy converted into photons and neutrinos. The energy of 
blue bump photons is taken to be 2.26 eV (V-band optical range). 
 Cosmological parameters used for calculation of luminosity distance are $H_0=70$
 km/s/Mpc, $\Omega_V=0.7,~ \Omega_M=0.3$. Points show the V-band optical flux from the sources
 listed in Tables \ref{tab:tev},  \ref{tab:jet} and \ref{tab:sources}.}
\label{fig:L_Dl}
\end{figure}

Blazars loud in the TeV energy range are often named as good candidate sources of neutrinos, because
the existence of TeV gamma-rays favors the possibility of proton acceleration up to high energies.
However, in order to produce neutrinos the  
acceleration of protons to energies above the pion production threshold is 
required, but not enough. A second important condition is the large optical depth 
for protons, Eq.~(\ref{bound}). Blazars from the Table \ref{tab:tev} are shown by stars 
(confirmed TeV sources) and by 
squares (possible TeV sources) in Fig.~\ref{fig:L_Dl}. All of 
them are marked with the same numbers 1-8, as in Table \ref{tab:tev}. One can see, that 
sources 1-6 do not obey the restriction Eq.~(\ref{bound}) and thus their neutrino flux 
is suppressed as compared to the photon flux.

Of course, the bound Eq.~(\ref{bound}) is only an order of magnitude argument. In order to
 estimate correctly the neutrino flux from a given source one needs to present a detailed model for this
source, explaining all experimental data. However, we can significantly improve the bound on the neutrino flux,
if we take into account   the fact that the source emits TeV gamma-rays. 
Let us make a simplest estimate of
the optical depth in the direction from the center of the core toward the Earth  
for TeV-blazars. The fact that TeV $\gamma$-rays produced in the vicinity of the central black hole 
are able to escape from the core and reach the Earth means that the mean free path of TeV photons with respect 
to  pair production on background photons,
\beq
R_\gamma=\frac{1}{\sigma_{\gamma\gamma}n_{\rm soft}}~,
\eeq
($\sigma_{\gamma\gamma}\sim \sigma_T=6.6\times 10^{-25}$ cm$^{2}$ is the cross-section of pair 
production for center-of-mass energies close to the pair production threshold;  
$n_{\rm soft}$ is the angle-dependent number density of the soft photons) is larger than the 
core size in the direction toward the Earth 
\beq
R_\gamma> R_{\rm core}.
\eeq  
The cross-section $\sigma_{p\gamma}\sim 10^{-28}$ cm$^{2}$ 
of interactions of protons with the same soft photons is more then on three orders of magnitude smaller than 
$\sigma_{\gamma\gamma}$. Thus, the mean free path for protons must be at least 
\beq
R_p\ge 10^3R_{\rm core}~,
\eeq
which means that just a negligible fraction of protons propagating in the direction of the interest interacts 
with the 
soft photons in the core. 
Thus, 
AGNs which are the sources of TeV $\gamma$-rays, like Mkn421, Mkn 501 or  1ES1426+428, can not be 
strong neutrino sources, since the proton energy can 
not be effectively converted into the energy of neutrinos. 
This, of course, does not mean that protons can not be accelerated in TeV blazars to ultra-high energies.
Observed TeV gamma-ray can be a result of synchrotron radiation from ultra-high energy protons \cite{aharonian2000}.

Of course, one can try to construct a model in which "hidden luminosity" 
in protons exceeds the observed TeV flux by three orders of magnitude. 
This would cause a problem of explaining the enormous energy balance of the source.

 Among the possible TeV sources 4-8 in Table \ref{tab:tev} only the first three do not 
obey the restriction 
Eq.~(\ref{bound}) and thus are similar to the confirmed TeV sources. 
However, PKS 2155-304 and 3C 66A  obey the bound Eq.~(\ref{bound})
and can be sources of neutrinos (if they are not real TeV sources).
Let us also note, that the HBL Lac  W Comae (1219+285) marked by number 9 in  Fig.~\ref{fig:L_Dl}
neither can  be a neutrino source according to Eq.~\ref{bound}. It is interesting to note that this 
source was recently suggested 
as a possible candidate for future TeV detection after the 
detailed analysis both in the synchrotron-self-Compton model
and in the proton blazar model \cite{WComae}.

\section{Most favorite neutrino sources.} 
\label{sec:four}

\begin{table}
\begin{center}
\begin{tabular}{|c|l|l|l|l|l|l|}
\hline
N & NAME   &z   &V mag  & Type & $F_{\rm GeV}$& Jet length\\
\hline
10& 3C 279  & 0.536 & 17.75 & HPQ &  $6.9 \pm 0.7$ & 14 kpc   \\
\hline
11& 4C 29.45  & 0.729 & 15.60 & HPQ &$1.9\pm 0.5$ &  16 kpc   \\
\hline
12& 3C 454.3  & 0.859 & 16.10 & HPQ &$3.5\pm 0.8$ & 21 kpc   \\
\hline
\hline
\end{tabular}
\caption{GeV loud blazars, which have large scale jets. 
Number in the first column same as in Fig.~\ref{fig:L_Dl}.
Second column is the name of the object.  Third is redshift. Forth is magnitude in V range
(eV range). Fifth is object type, six is GeV flux in units of $10^{-8}$ s$^{-1}$cm$^{-2}$  and the last is jet length in kpc.} 
\label{tab:jet}                             
\end{center}
\end{table}

\begin{table*}
\caption{Blazars which satisfy the selection criteria {\bf 1-3}. 
Coordinates are in Equatorial J2000 system.
GeV flux in units of $10^{-8}$ s$^{-1}$cm$^{-2}$.}
\label{tab:sources}                             
\begin{tabular}{|c|l|l|l|l|l|l|}
\hline
NAME     &  Longitude     &    Latitude  &   z   &  V mag   &    $F_{\rm GeV}$&   Type\\
\hline
QSO 0208-512~~~  & 32.58~~~~  &-50.93~~ ~~ & 1.003~~ ~~ & $16.9$~~~~ &  $8.5\pm 1.2$~~ ~~~  & HPQ~~~~ \\
\hline
QSO 0219+428  & 35.70  &  42.90 & 0.444  & $15.5$ &   $2.8\pm 0.7$  &  LBL\\
\hline
QSO 0235+164  &  39.36 &  16.59 & 0.940  & $15.5$ &   $5.5\pm 1.2$  &  LBL\\
\hline
QSO 0440-003  &70.55   &  -0.55 & 0.844 & $19.2$&   $1.4\pm 0.5$    &  HPQ\\
\hline
QSO 0528+134  & 82.74  & 13.38  & 2.060 & $20.0$&   $3.0\pm 0.5$    &   LPQ\\
\hline
QSO 0537-441  & 85.02  & -44.05 & 0.894  & $15.5$&   $2.3\pm0.7$    &   LBL\\
\hline
QSO 0716+714  & 110.47  & 71.34& 0.3  & $14.17$&   $1.9\pm 0.5$    &   LBL\\
\hline
QSO 0954+556  & 148.01 & 55.02  & 0.901  & $17.7$&   $1.4\pm 0.4$   &  HPQ \\
\hline
QSO 1406-076 & 212.42 &  -7.75 & 1.494  & $18.4$&   $2.0\pm 0.6$    &  LPQ\\
\hline
QSO 1611+343 & 243.54 & 34.40  & 1.401& $17.5$&   $2.3\pm 0.8$      & LPQ \\
\hline
QSO 1633+382 & 248.92 & 38.22  & 1.814 & $18.0$&   $4.8\pm 1.1$     &  LPQ \\
\hline
QSO 1730-130  & 263.46 & -13.23 & 0.902   & $18.5$&   $2.4\pm 0.6$ & FSRQ \\
\hline
QSO 2005-489  & 302.4  & -48.8  & 0.071  & $13.4$&   $2.2\pm 0.8$ &   FSRQ \\
\hline
QSO 2022-077 & 306.36 &  -7.75 & 1.388   & $18.5$&   $2.5\pm 0.8$ &  FSRQ \\
\hline
QSO 2155-304 & 329.0 &  -30.5  & 0.116   & $13.1$&   $<1.5$ &  HBL \\
\hline
\end{tabular}
\end{table*}

If the photon emission in the $10^{14-16}$ eV range is highly beamed, 
one would expect that the neutrino emission in the 
same energy range is also highly beamed since both photons and neutrinos 
are presumably produced through the 
photo-pion process  by the protons accelerated in the AGN core. The assumption about an 
anisotropic character of the neutrino emission can change dramatically 
predictions about possible detection of neutrino point sources in neutrino telescopes. 
For example, some $\gamma$-ray loud blazars, like 3C 273 or 3C 279 which were conjectured to be 
strong neutrino emitters \cite{3C273} would not be detected by neutrino telescopes because their 
jets are not oriented  along the line of sight (the jet in 3C273 has projected size about 
39 kpc, while the one in 3C279 is 14 kpc long \cite{jets}). Moreover, blazar 3C 273  was not even included in our 
list of sources, because its $\gamma$-ray flux in the GeV energy range  is small.

The above argumentation enables one to work out simple selection criteria for possible neutrino
sources:
\begin{itemize}
\item {\bf 1} AGN       is $\gamma$-ray loud in GeV range. 
\item {\bf 2} The AGN  luminosity satisfies the  bound Eq.~(\ref{bound}).
\item {\bf 3} A large scale jet is either not observed or its length is less than 1 kpc.
\end{itemize}
(The last condition roughly insures that the AGN is seen at a viewing 
angle $\theta\le 1^\circ$ if we suppose that 
the typical length of the  the large scale jet is 100 kpc.)

 From our list of sources (27 objects), which obey the condition {\bf 1}, we exclude 
 9 which do not obey the bound  Eq.~(\ref{bound}). Then we have  separated 3 objects in which the 
large scale (with length more than 1 kpc) jets are 
detected using the catalog of extragalactic jets \cite{jets}. Those objects are listed in Table \ref{tab:jet} and presented in Fig.~\ref{fig:L_Dl} with the numbers 10-12.
 The 15 sources left after the selection   
procedure are listed in the Table \ref{tab:sources}.

\begin{figure}
\includegraphics[width=0.6\linewidth,clip=true,angle=270]{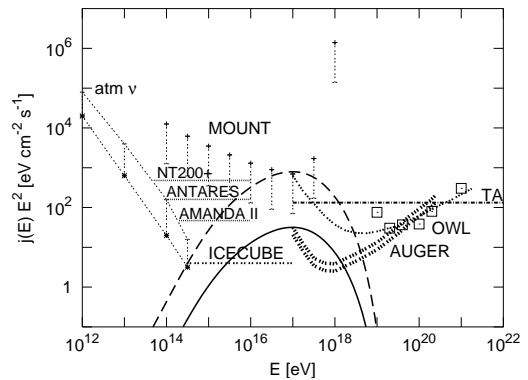}
\caption{Neutrino flux from typical GeV-loud blazar from Table \ref{tab:sources} (thick solid line)
 compared with expected sensitivities to electron/muon and tau-neutrinos~
in detectors AMANDA II \cite{amanda_limit}, Auger \protect\cite{auger_nu},
and the planned projects: Telescope Array (TA)~\protect\cite{ta_nu} (dashed-dotted line), the
fluorescence/\v{C}erenkov detector MOUNT~\protect\cite{mount}, the
space based OWL~\protect\cite{owl_nu} (indicated by squares) (we take the latter as representative
also for EUSO), the water-based NT200+~\protect\cite{baikal_limit},
ANTARES~\protect\cite{antares} (the NESTOR~\protect\cite{nestor} sensitivity would
be similar to ANTARES according to Ref.~\protect\cite{nu_tele}), and
the ice-based ICECUBE~\protect\cite{icecube}, as indicated.  All not published experimental sensitivities 
 are scaled from corresponding diffuse sensitivities with  the same factor
as ICECUBE.  
The dashed line is for an opening angle for neutrino 5 times smaller than the openinning angle for GeV photons. }    
\label{fig:detect_angle}
\end{figure}

Assuming that the GeV $\gamma$-ray flux from these sources comes mostly from the cascaded photons produced
in the photo-pion process we can estimate the neutrino flux from each source, using the fact that the 
energy deposit in photons is of the same order as the energy deposit in neutrinos in photo-pion production
process. 

However, it is important to note that  the $\gamma$-ray flux from a given source can be even lower than 
the neutrino flux due to a variety of reasons. First of all, if the proton optical depth is not 
too high, say $\tau_p\sim 10$,  VHE $\gamma$-rays with energies $E_\gamma>10^{17}$ eV partially 
escape from the
AGN core and dissipate their energy in the large scale jet. Thus, the power carried by the  photon beam  
is transmitted to the 100 kpc-scale jet \cite{gamma_jet}. Next, if there is a strong 
(disordered) magnetic field  inside the core, 
pairs produced by VHE photons can lose their energy mostly on  synchrotron radiation rather than on inverse 
Compton scattering  
of ambient photons. In this case the power contained in the photon beam will mostly go into synchrotron 
photons with the energies below MeV rather than to GeV-TeV ranges. 
Finally,
several sources which have jets  seen face-on can be very bright neutrino sources 
with the neutrino flux much larger then the observed GeV flux.
 Indeed, in the photo-pion process the 
total energy emitted in $10^{14-17}$ eV photons and neutrinos are of the same order. But the neutrino 
flux remains collimated within $1^\circ$ all over the propagation distance to the Earth, while 
the $\gamma$-ray flux looses its collimation during the development of electromagnetic  cascades
on the soft radiation background in the AGN core and in the intergalactic medium. The cascade ends in the GeV
energy range and the GeV $\gamma$-ray flux from a blazar is emitted into a cone with larger opening angle,
as it is shown in Fig. \ref{fig:cartoon}.  

In Fig. ~\ref{fig:detect_angle} we presented neutrino flux from GeV-loud blazar in two cases:
when the neutrino flux is similar to photon flux, and when neutrino flux is collimated
in small angle (1 degree instead of 5 degree for GeV photons). In first case only ICECUBE and 
Piere Auger Observatory will be able to detect neutrino fluxes from point-like sources.
In the last case many other experiments will be able to see the neutrino flux from the sources in the Table \ref{tab:sources}.
 However, the smaller opening angle for neutrino flux will reduce the number of neutrino sources.

\section{Ultra-high energy neutrinos from blazars.}
\label{sec:five}

In the previous Section we have considered the case when protons are accelerated 
up to the energies $10^{18} -10^{19}$ eV. However,  UHECR with energies up to 
$3 \times 10^{20}$eV \cite{Fly} were 
observed.
This means that in principle the sources of UHECR can accelerate protons at least up to energies  $E\sim  10^{21}$ eV.
Let us note here, that the recent disagreement in results between AGASA \cite{AGASA} and HiRes \cite{HiRes} experiments does not
raise the question of  the existence or non-existence of events with energy $E > 10^{20}$eV. Both experiments
detect events at those energies and there is disagreement only in the number of such events. 

It is interesting to note that the bound Eq.~(\ref{bound}) on the magnitude-redshift of blazars which can be 
neutrino-loud can be converted into a bound on UHECR-loud sources, if we just turn $\gg$ into $\ll$. Indeed,
if we suppose that ultra-high energy protons and photons are able to leave the core of AGN the optical depth 
for them must be extremally low. In this respect we note that the set of EGRET blazars 
presented in Table \ref{tab:sources} is "anti-correlated" with the selection of EGRET blazars whose positions coincide
 with the arrival directions of UHECR presented in \cite{tkachev2002}.

\begin{figure}
\includegraphics[width=0.6\linewidth,clip=true,angle=270]{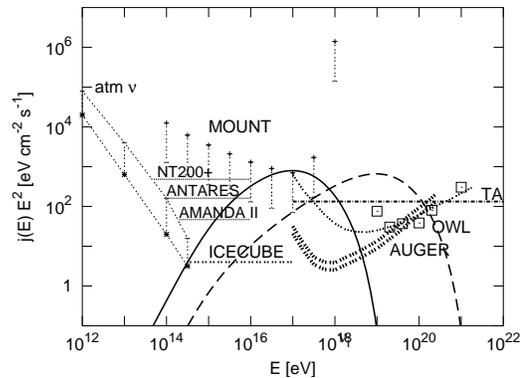}
\caption{Same neutrino flux as in Fig.~\ref{fig:detect_angle}   for 1 degree opening angle
in case $E_p \sim 10^{19}$ eV (solid line) and neutrino flux from the same source in case $E_p\sim  10^{21}$ eV (dashed line).
Fluxes compared to sensitivities of future detectors. Line keys for experiments are the same as in 
Fig.~\ref{fig:detect_angle}. }    
\label{fig:detect_energy}
\end{figure}

However, blazars with large optical depth for protons still could be  significant sources of UHE neutrinos
if we assume that protons are accelerated in the source up to energies $10^{20}-10^{21}$ eV.
The resulting neutrino spectrum will be very different
from the one discussed in the previous Section. We presented both spectra in Fig.~ \ref{fig:detect_energy}.
The neutrino flux from blazars accelerating protons up to the highest energies  will be   
peaked in the region  $10^{19}$ eV or above and will be  seen by future UHECR experiments,
while neutrino flux from "moderate accelerators" of protons will be peaked at the 
energies  $10^{16} - 10^{17}$ eV 
and can be detected by future neutrino telescopes.
Let us note, that constraints on neutrino sources, discussed recently in \cite{nu_source},
are not applicable to blazars, which produce neutrinos in the AGN cores.

As it is argued at the end of  Section \ref{sec:four}, the detected $\gamma$-ray flux from 
a given blazar can serve only as 
a "hint", not as the upper limit on the neutrino flux. By the same arguments this is also true for the case of 
flux of UHE neutrinos. Blazars in which 
the neutrino flux is peaked at high energies as in Fig.~\ref{fig:detect_energy} 
and is collimated in smaller opening angle as compared to the photon flux can serve as 
 "pure neutrino sources", which are required for the so-called 
Z-burst model. In the Z-burst scenario the UHECRs are produced by Z-bosons
decaying within the distance relevant for the Greisen-Zatsepin-Kuzmin (GZK) effect \cite{gzk}. These
Z-bosons are in turn produced by UHE neutrinos interacting with
the relic neutrino background~\cite{zburst1} (for recent detailed numerical simulations see \cite{kkss,ringwald}). 
The flux of photons from astrophysical 
sources in this model should be significantly suppressed in comparison to the neutrino flux, otherwise the Z-burst
model would be ruled out, as demonstrated
in Ref.~\cite{kkss}. The spectrum presented in Fig.~\ref{fig:detect_energy} can serve as a
prototype for the spectrum from a "pure neutrino source". However, in Z-burst model 
the protons should be accelerated even to higher energies $E > 10^{23}$~eV and 
the UHECRs with highest energies $E> 10^{20}$~eV should come from few strong neutrino sources.
The last condition is required to overcome bound on diffuse neutrino flux \cite{diffuse}, which come 
from diffuse gamma-ray flux, measured by EGRET\cite{egret}.

\section{Conclusions}

In this work we have considered conditions under which $\gamma$-ray loud blazars can be significant 
neutrino sources. High energy neutrinos are produced in the photo-pion interactions of protons accelerated
in the AGN cores with soft photon background. We have derived the bound Eq.~(\ref{bound}) on 
the redshift and V-magnitude of the candidate neutrino-loud blazars. 
From 27 GeV-TeV $\gamma$-ray loud blazars we selected 15 most favorite candidates which satisfy 
the criteria {\bf 1-3} listed in Section \ref{sec:four}.

An estimate of the neutrino flux from a given object can be obtained from the observed $\gamma$-ray flux if 
we assume that the main contribution to the GeV-TeV luminosity of a blazar comes from the electromagnetic
cascade initiated by $10^{14-16}$ eV photons which are produced together with neutrinos in photo-pion reactions.
It is important to note that the neutrino flux from a given source  can be much higher than the $\gamma$-ray flux 
detected by EGRET or other $\gamma$-ray telescopes due to the variety of reasons listed in Section \ref{sec:four}.

Because the optical depth for TeV photons in the AGN core 
is three orders of magnitude smaller than the  optical
depth for protons in the same photon background we conclude that confirmed TeV-loud blazars
can not be  sources of significant neutrino flux. (At the same time, 
this argument does not exclude the possibility that these objects can be  neutrino sources  
from UHECR protons which produce neutrinos in interactions with CMBR photons.)   

We have considered the  model in which the neutrino flux is highly beamed in the directions of the large scale 
jets emitted by the AGN. This model has several experimentally testable predictions.
 First, GeV-loud sources  in which the large scale jets are not seen face on (like 3C 279)  
can not be neutrino sources. Next, the neutrino flux from a given source can be much larger than the 
observed gamma-ray flux
due to a smaller opening angle for neutrinos as compared to GeV gamma-rays.  

 We have also found that if the optical depth for protons is large enough and protons are 
accelerated up to the highest energies, 
$10^{20}-10^{21}$ eV, the same sources from Table \ref{tab:sources} can be seen both by future UHECR detectors  and
by neutrino telescopes, see  Fig.~\ref{fig:detect_energy}. 

We conclude that the next generation  of neutrino telescopes and UHECR detectors will have a good chance to 
see  point-like neutrino sources.

\section*{Acknowledgments}
We would like to thank Felix Aharonian, Michael Kachelriess, Guenter Sigl and Igor Tkachev for
fruitful discussions and comments.

\end{document}